\documentclass[review]{elsarticle}

\makeatletter
\def\ps@pprintTitle{
	\let\@oddhead\@empty
	\let\@evenhead\@empty
	\def\@oddfoot{\reset@font\hfil\thepage\hfil}
	\let\@evenfoot\@oddfoot
}
\makeatother

 \biboptions{comma,sort&compress}

\usepackage{graphicx}
\usepackage{amsmath}
\usepackage{here}
\usepackage{natbib}
\usepackage{cuted}
\usepackage{dashrule}
\usepackage[usenames,dvipsnames]{color}
\usepackage{xcolor}

\usepackage[dvips]{epsfig}

\newcommand{\clabel}[2][]{#2}
\newcommand{\change}[1]{#1}
\begin{document}



\markboth{Wei-Lin Wu, Peking University (CN)}{ }
\begin{frontmatter}

\title{Tetraquark states in the quark model\,\tnoteref{invit}}
\tnotetext[invit]{Talk given at QCD25 - 40th anniversary of the QCD-Montpellier Conference. }
\author{Wei-Lin Wu}
\address{School of Physics, Peking University, Beijing 100871, China}
\ead{wlwu@pku.edu.cn}
\author{Yan-Ke Chen}
\author{Yao Ma}
\address{School of Physics and Center of High Energy Physics,
	Peking University, Beijing 100871, China}
\author{Lu Meng}
\address{School of Physics, Southeast University, Nanjing 210094, China}
\author{Shi-Lin Zhu}
\address{School of Physics and Center of High Energy Physics,
	Peking University, Beijing 100871, China}


\date{\today}
\begin{abstract}
We perform systematical investigations of heavy flavor tetraquark systems, including fully heavy $QQ\bar Q\bar Q$ ($Q=b,c$), doubly heavy $QQ^{(\prime)}\bar q\bar q$ ($q=u,d$), and singly heavy $Qs\bar q\bar q$ tetraquark systems, within the framework of quark potential model. We employ the Gaussian expansion method and complex scaling method to solve the four-body Hamiltonian and identify bound and resonant states. We further calculate the root mean square radii to study the spatial configurations of tetraquarks. Our calculations reveal a rich spectrum of tetraquark states exhibiting diverse spatial configurations. In particular, we find good candidates for experimental states $X(6900)$, $X(7200)$, $T_{cc}(3875)^+$, and $T_{\bar c\bar s0}^*(2870)$. The fully charmed tetraquark resonances $X(6900)$ and $X(7200)$ are compact teraquark states, while doubly charmed tetraquark bound state $T_{cc}(3875)^+$ and charm-strange tetraquark resonance $T_{\bar c\bar s0}^*(2870)$ are meson molecules. Additionally, more tetraquark bound and resonant states are predicted and may be searched for in future experiments.

\end{abstract}
\end{frontmatter}
\newpage
\section{Introduction}
In recent decades, numerous exotic hadrons beyond the conventional meson and baryon schemes in the simple quark model have been observed experimentally. For some of these states, the multiquark content is manifestly revealed by their quantum numbers or masses, such as the doubly charmed tetraquark $T_{cc}(3875)^+\,(cc\bar u\bar d)$~\cite{LHCb:2021vvq,LHCb:2021auc}, charm-strange tetraquarks $T^*_{\bar c\bar s0}(2870)^0,T^*_{\bar c\bar s1}(2900)^0\,(\bar c\bar s u d)$~\cite{LHCb:2020bls,LHCb:2020pxc}, and fully charmed tetraquarks $(cc\bar c\bar c)$~\cite{LHCb:2020bwg,ATLAS:2023bft,CMS:2023owd}. Investigating the binding mechanisms and clustering behaviors of these tetraquark states may help advance our understanding of the nonperturbative quantum chromodynamics (QCD).

Extending the quark potential model to multiquark configurations may serve as an approach to understanding the exotic hadrons in a unfied framework. The potential parameters used to calculate the tetraquark spectrum can be determined from the known meson or baryon spectra. Since no extra fits to the experimental tetraquark results are needed, the quark model can provide forward-looking predictions of various tetraquark systems. Additionally, the quark model makes no \textit{a priori} assumptions about the clustering of quarks. Therefore, it can consistently describe tetraquark states with different configurations, including meson molecules and compact tetraquarks. \clabel[q3]{Previous works used the quark potential model to study doubly heavy tetraquark bound states~\cite{Meng:2020knc,Deng:2021gnb}. Doubly heavy tetraquark resonances were also explored using the real scaling method~\cite{Meng:2021yjr,Meng:2023for}. Ref.~\cite{Wang:2022yes} studied fully charmed tetraquark resonances using the complex scaling method. In this work, we perform systematical studies of various heavy flavor tetraquark systems, including fully heavy tetraquark $(cc\bar c\bar c, bb\bar b\bar b, bc\bar b\bar c)$, doubly heavy tetraquark $(cc\bar q\bar q, bb\bar q\bar q, bc\bar q\bar q)$ and singly heavy tetraquark $(cs\bar q\bar q,bs\bar q\bar q)$, where $q=u,d$.   We utilize effective few-body methods, including the Gaussian expansion method~\cite{HIYAMA2003223} and complex scaling method~\cite{Aguilar1971,Balslev1971,Aoyama2006}, to solve the Schr\"odinger equation and identify bound and resonant states. We further identify the spatial structures of tetraquark states by calculating their root-mean-square radii. We find candidates for a series of experimental states and provide predictions for future states.}
\section{Formalism}
\subsection{Hamiltonian}
In the nonrelativistic quark model framework, the dynamics of tetraquark systems is described by the Hamiltonian,
\begin{equation}
	H=\sum_{i=1}^4 (m_i+\frac{p_i^2}{2 m_i})+\sum_{i<j=1}^4 V_{ij
	}-T_{C.M.},
\end{equation}
where $m_i$ and $p_i$ are the mass and momentum of (anti)quark $i$, $V_{ij}$ denotes the two-body interaction, and the center-of-mass kinetic energy $T_{C.M.}$ is subtracted. The interaction $V_{ij}$ takes various forms in different models. In this work, we adopt three models which contain one-gluon-exchange and quark confinement terms, including the AL1 and AP1 model proposed in Ref.~\cite{SilvestreBrac1996} and the BGS model proposed in Ref.~\cite{Barnes:2005pb}. In the S-wave systems, the potential is written as
\begin{equation}
	\label{eq:potential}
		V_{i j} =-\frac{3}{16} \boldsymbol\lambda_i \cdot \boldsymbol\lambda_j\left(-\frac{\kappa}{r_{i j}}-\Lambda+\frac{8 \pi \kappa^{\prime}}{3 m_i m_j} \frac{\exp \left(-r_{i j}^2 / r_0^2\right)}{\pi^{3 / 2} r_0^3} \boldsymbol{S}_i \cdot \boldsymbol{S}_j+\lambda r_{i j}^p\right).
\end{equation}  
Here, $\boldsymbol\lambda_i$ (replaced by  $-\boldsymbol\lambda_i^*$ for antiquark) and $\boldsymbol{S}_i$ are the SU(3) color Gell-Mann matrix and spin operator of (anti)quark $i$. The AL1 and BGS model contains a linear confinment term ($p=1$), while the AP1 model contains a $p=\frac{2}{3}$-power confinement term. The parameters in the models were all fitted by the meson spectra and can be found in Refs.~\cite{SilvestreBrac1996,Barnes:2005pb}. The theoretical masses of the mesons are compared with the experimental ones in Table~\ref{tab:meson}. 
Results from all three models agree with the experimental values within tens of MeV. We expect the model uncertainties of tetraquark states to be of the same order.

\begin{table}[hbt]
	\begin{center}
		\setlength{\tabcolsep}{0.5pc}
		
		\caption{\footnotesize    
			Meson mass spectra (in MeV) from the AL1, AP1 and BGS models. The experimental values taken from Ref.~\cite{ParticleDataGroup:2024cfk} are listed for comparison.}
		{\footnotesize
			\begin{tabular}{llllllllll}
				\hline\hline
				Mesons&$K$&$D$&$D_s$&$B$&$B_s$&$\eta_c$&$\eta_c(2S)$&$\eta_b$&$\eta_b(2S)$\\
				\hline 
				Exp.&494&1867&1968&5279&5367&2984&3638&9399&9999\\
				AL1 &491&1862&1965&5293&5362&3006&3608&9424&10003\\
				AP1 &498&1881&1955&5311&5356&2982&3605&9401&10000\\
				BGS &-&-&-&-&-&2982&3630&-&-\\
				\hline
				Mesons&$K^*$&$D^*$&$D_s^*$&$B^*$&$B_s^*$&$J/\Psi$&$\Psi(2S)$&$\Upsilon$&$\Upsilon(2S)$\\
				\hline 
				Exp.&895&2009&2112&5325&5415&3097&3686&9460&10023\\
				AL1 &904&2016&2102&5350&5418&3101&3641&9462&10012\\
				AP1 &908&2033&2107&5367&5418&3102&3645&9461&10014\\
				BGS &-&-&-&-&-&3090&3672&-&-\\
				\hline\hline
			\end{tabular}
		}
		\label{tab:meson}
	\end{center}
\end{table}
\subsection{Complex Scaling Method}
Resonant states are poles of the S matrix on the complex energy plane. The resonant wave functions are not square integrable, causing numerical difficulties for obtaining resonances. The complex scaling method~\cite{Aguilar1971,Balslev1971,Aoyama2006} enables solving for tetraquark bound states and resonant states simultaneouly by carrying out analytical continuation of the Hamiltonian. The coordinate $\mathbf{r}$ and its conjugate momentum $\mathbf{p}$ are transformed as
\begin{equation}
	U(\theta) \boldsymbol{r}=\boldsymbol{r} e^{i \theta}, \quad U(\theta) \boldsymbol{p}=\boldsymbol{p} e^{-i \theta},
\end{equation}
where $\theta$ is the complex-scaling angle. The complex-scaled Hamiltonian reads
\begin{equation}
	H(\theta)=\sum_{i=1}^4 (m_i+\frac{p_i^2e^{-2i\theta}}{2 m_i})+\sum_{i<j=1}^4 V_{ij}(r_{ij}e^{i\theta}).
\end{equation}
The bound states, resonant states and continuum states can be obtained simultaneously by solving the energy spectrum of the complex-scaled Hamiltonian. Bound states lie on the real axis below thresholds. Continuum states align along rays originating at threshold energies with $\operatorname{Arg}E=-2\theta$. Resonant states appear at $E_R=M_R-i\Gamma_R/2$ when $2\theta>|\operatorname{Arg}E_R|$, where $M_R$ and $\Gamma_R$ are the mass and width of the resonance. 
\subsection{Gaussian Expansion Method}
\begin{figure}[hbt]
	\centering
	\includegraphics[width=8cm]{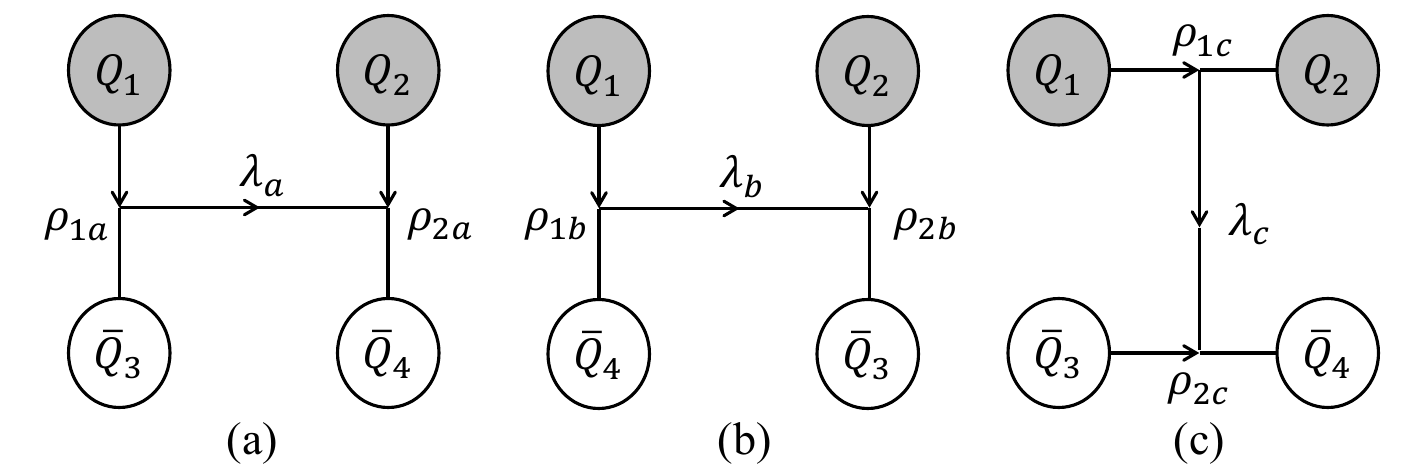}
	\caption{\scriptsize The Jacobian coordinates for two types of spatial configurations: (a), (b) for the dimeson configurations, and (c) for the diquark-antidiquark configuration.}
	\label{fig:jac}
\end{figure}

The Gaussian expansion method is shown to be an effective numerical approach for solving few-body Schr\"odinger equation~\cite{HIYAMA2003223}. The four-body spatial wave function is expanded by the Gaussian basis functions,
\begin{equation}\label{eq:spatial_wf}
		\Phi^{L}_{n_{1},n_{2},n_{3}} = \,\phi_{n_1l_{\rho_1}}(\rho_{1})\phi_{n_{2}l_{\rho_2}}(\rho_{2})\phi_{n_{3}l_\lambda}(\lambda)\left[\left(Y_{l_{\rho_1}}(\hat{\rho_1})\otimes Y_{l_{\rho_2}}(\hat{\rho_2})\right)\otimes Y_{l_\lambda}(\hat{\lambda})\right]^{L},
\end{equation}
where $L$ denotes the total orbital angular momentum,  $\rho_1,\rho_2,\lambda$ are three independent Jacobian coordinates in three types of configurations considered in this work, as shown in Fig.~\ref{fig:jac}. $ \phi_{nl}(r) $ takes the Gaussian form,
\begin{equation}
	\begin{array}{c}
		\phi_{nl}(r)\propto r^le^{-\nu_{n}r^2},\\
		\nu_{n}=\nu_{1}\gamma^{n-1}\quad (n=1\sim n_{\rm max}).
	\end{array}
\end{equation}

The incorporation of dimeson and diquark-antidiquark configurations enables descriptions of both molecular and compact states. The Gaussian basis functions can well describe both short-range and long-range correlations~\cite{HIYAMA2003223}. In addition, the matrix elements of Gaussian bases can be solved analytically~\cite{HIYAMA2003223,Wu:2024ocq}, making large-scale basis expansion calculations numerically feasible. The complete four-body wave function can be written as,
\begin{equation}\label{eq:wavefunction}
	\begin{aligned}
		\Psi^{J}(\theta)=\mathcal{A}\sum_{\rm{(jac)}}\sum_{\alpha,n_{i}}\,C^{(\rm{jac})}_{\alpha,n_{i}}(\theta)\left[\chi_\alpha^S\otimes\Phi^{L,(\rm{jac})}_{n_{1},n_{2},n_{3}}\right]^J,
	\end{aligned}
\end{equation}
where \clabel[q1]{$J$ is the total angular momentum, $ \mathcal{A} $ is the antisymmetric operator of identical particles, $C$ is the expanded coefficients. $\rm{(jac)},\alpha,$ and $n_i$ sum over the three types of Jacobian configurations in Fig.~\ref{fig:jac}, the color-spin configurations, and the radial indices of Gaussian bases, respectively. } $ \chi_\alpha^S $ is the color-spin wave function, given by
\begin{equation}\label{eq:colorspin_wf1}
	\begin{aligned}
		\chi^{S}_{\bar 3_c\otimes 3_c,s_1,s_2}=\left[\left(Q_1Q_2\right)_{\bar 3_c}^{s_1}\left(\bar Q_3\bar Q_4\right)_{3_c}^{s_2}\right]_{1_c}^{S},\\
		\chi^{S}_{6_c\otimes \bar 6_c,s_1,s_2}=\left[\left(Q_1Q_2\right)_{6_c}^{s_1}\left(\bar Q_3\bar Q_4\right)_{\bar{6}_c}^{s_2}\right]_{1_c}^{S},\\
	\end{aligned}
\end{equation}
for all possible spin combinations $ s_1, s_2, S $. \clabel[q2]{It is worth noting that the  diquark-antidiquark type color-spin basis in Eq.~\eqref{eq:colorspin_wf1} employed in this work is equivalent to the dimeson type color-spin basis,
	\begin{equation}\label{eq:colorspin_wf2}
		\begin{aligned}
			\chi^S_{1_c\otimes 1_c,s_1,s_2}=\left[\left(Q_1\bar Q_3\right)^{s_1}_{1_c}\left(Q_2\bar Q_4\right)^{s_2}_{1_c}\right]^S_{1_c},\\
			\chi^S_{8_c\otimes 8_c,s_1,s_2}=\left[\left(Q_1\bar Q_3\right)^{s_1}_{8_c}\left(Q_2\bar Q_4\right)^{s_2}_{8_c}\right]^S_{1_c}.\\
		\end{aligned}
	\end{equation}
They are both complete bases of the tetraquark color-spin degrees of freedom, and can be related to each other by unitary transformation~\cite{Wu:2024euj}.}
\subsection{Spatial structures}
\clabel[q5]{To investigate the clustering behaviors of tetraquark states, we use the inter-quark root-mean-square (rms) radii to distinguish between meson molecules and compact tetraquarks. In a meson molecule, the four (anti)quarks cluster into two color-singlet mesons, with their relative distances being larger than the typical color confinement scale $\Lambda_{QCD}^{-1}\sim 1\,\text{fm}$. In a compact tetraquark, the four (anti)quarks are confined together within the scale of  $\Lambda_{QCD}^{-1}$. The inter-quark rms radii can be used to reveal different spatial structures. However, in the presence of identical particles, the conventional rms radii calculated using the complete wave function can be ambiguous and fail to reveal molecular configuration due to the antisymmetrization ~\cite{Chen:2023syh}. We propose to calculate the rms radii as
\begin{equation}\label{eq:rmsr}
		r_{ij}\equiv \mathrm{Re}\left[\sqrt{\frac{\langle\Psi_{\mathrm{nA}}(\theta) | r_{ij}^2 e^{2i\theta}|\Psi_{\mathrm{nA}}(\theta)\rangle}{\langle\Psi_{\mathrm{nA}}(\theta) | \Psi_{\mathrm{nA}}(\theta)\rangle}}\right].
\end{equation}
Here, $ij$ are the quark indices, $\theta$ is the complex scaling angle, and $\Psi_{\mathrm{nA}}$ is the decomposed nonantisymmetric wave function, where $Q_1\bar Q_3$ and $Q_2\bar Q_4$ form two color singlets~\cite{Wu:2024zbx}. In the molecular scheme, $r_{13}$ and $r_{24}$ describe the sizes of constituent mesons, while $r_{12},r_{14},r_{23},$ and $r_{34}$ reflect the inter-meson distance. In the compact tetraquark scheme, the rms radii defined by Eq.~\eqref{eq:rmsr} are consistent with the conventional definition.  More discussions on the rms radii can be found in Ref.~\cite{Wu:2024zbx}.}

\section{Results}
\subsection{Fully heavy tetraquark}
We study the S-wave fully charmed tetraquark systems using three different potential models, including the AL1, AP1 and BGS model. We find a series of resonant states, whose energies are listed in Table~\ref{tab:4c_energies}~\cite{Wu:2024euj}. The three models give qualitatively consistent energy spectra: Most resonances exist in all three models with similar masses and widths.  Moreover, the tetraquark spectra in systems with different quantum numbers $J^{PC}$ demonstrates analogous characteristics. A lower state at $M\approx 7$ GeV with width $\Gamma\approx75$ MeV, and a higher state at $M\approx 7.2$ GeV with width $\Gamma \approx 50$ MeV are found in the $0^{++},1^{+-},2^{++}$ systems. The lower $0^{++}$ and $2^{++}$ states are good candidates for the experimental $X(6900)$ state~\cite{LHCb:2020bwg,ATLAS:2023bft,CMS:2023owd}, while the higher $0^{++}$ and $2^{++}$ states are good candidates for the experimental $X(7200)$ state~\cite{ATLAS:2023bft,CMS:2023owd}. Both the masses and widths agree with the experimental values within tens of MeV. The similar pattern of spectra with different $J^{PC}$ is expected, as the spin splitting term in Eq.~\eqref{eq:potential} is suppressed by the heavy quark mass. However, recent analyses from the CMS collaborations suggest that the experimental states have $J^{PC}=2^{++}$~\cite{CMS:2025fpt}. Further studies are needed to better understand and resolve the discrepancy. 

\begin{table}[htbp]
	\centering
	\caption{\footnotesize The complex energies $ E=M-i\Gamma/2 $ (in $\mathrm{MeV}$) of the fully charmed tetraquark resonances from three potential models.  The  "?"  indicates that potential signs of resonant states are observed but the pole positions cannot be obtained accurately in the present calculations. The "-" suggests that no corresponding resonance is found.  \change{The last column shows possible experimental correspondences of the theoretical states.} }
	\label{tab:4c_energies}
	{\footnotesize \begin{tabular}{ccccc}
		\hline\hline
		$ J^{PC} $& AL1 & AP1& BGS & EXP\\
		\hline
		$ 0^{++} $&$ 6980-35i $&  $ 6978-36i	 $&$ 7030-36i $&$X(6900)$\\
		&$ 7034-1i $&$ 7049-1i $&$ 7127-0.1i $\\
		&$ 7156-20i $&$ 7167-19i $&$ 7239-17i $&$X(7200)$\\
		
		$ 1^{+-} $&$ 6921-0.5i $&$ 6932-0.5i $&$ 6991-0.1i $\\
		&$ 6995-35i $&$ 6998-35i $&$ 7048-35i $\\
		&?&$ 7181-27i $&$ 7254-24i $\\
		$ 2^{++} $&$ 7013-38i $&$ 7017-39i$&$ 7066-39i$&$X(6900)$\\
		&$ 7127-6i $&$ 7114-4i $&-\\
		&?&$ 7204-29i $&$ 7268-32i $&$X(7200)$\\
		& $ 7272-9i $ &$ 7276-12i $&$ 7337-8i $\\
		\hline\hline
	\end{tabular}}
\end{table} 

\clabel[q4]{As the spectra from the three models are similar, we choose the results from the AP1 model to study the resonant structures. The color proportions and the rms radii are shown in Table~\ref{tab:4c_structure}. From the rms radii, we find that all fully charmed tetraquark resonances exhibit a compact tetraquark configuration, with four (anti)quarks confined within the typical range of color confinement $\Lambda_{\text QCD}^{-1}\sim 1$ fm. This is expected as the interactions between heavy (anti)quarks should be dominated by short-range gluon exchange. The candidates for $X(6900)$ are dominant by the $\chi_{\bar 3_c\otimes 3_c}$ configuration $(\sim90\%)$, while the candidates for $X(7200)$ show strong mixing effects between $\chi_{\bar 3_c\otimes 3_c}$ and $\chi_{6_c\otimes\bar 6_c}$.}
\begin{table*}[htbp]
		\centering
		\caption{\footnotesize The proportions of different color configurations and the rms radii (in fm) of the $ cc\bar c\bar c $ resonant states in the AP1 potential. The last column shows the spatial configurations of the states, where C. represents the compact tetraquark. }
		\label{tab:4c_structure}
		{\footnotesize \begin{tabular}{ccccccccc}
			\hline\hline
			$ J^{PC} $&$  M-i\Gamma/2 $ &$ \chi_{\bar 3_c\otimes 3_c} $& $ \chi_{6_c\otimes\bar6_c} $& $ r_{c_1\bar{c}_3} $&$ r_{c_2\bar{c}_4} $&$ r_{c_1\bar{c}_4}$=$ r_{c_2\bar{c}_3} $&$ r_{c_1c_2} $=$ r_{\bar{c}_3\bar{c}_4} $& Configurations\\
			\hline
			
			$ 0^{++} $&$ 6978-36i $&$ 86\% $&$14\%$&$0.81$&$0.81$&$0.86$&$0.66$&C.\\
			&$ 7049-1i $&$37\%$&$63\%$&$0.70$&$0.70$&$0.82$&$0.75$&C.\\
			&$ 7167-19i $&$46\%$&$54\%$&$0.91$&$0.91$&$0.90$&$0.67$&C.\\
			$ 1^{+-} $&$ 6932-0.5i $&$65\%$&$35\%$&$0.66$&$0.66$&$0.73$&$0.63$&C.\\
			&$ 6998-35i $&$88\%$&$12\%$&$0.79$&$0.80$&$0.77$&$0.59$&C.\\
			&$ 7181-27i $&$44\%$&$56\%$&$0.91$&$0.93$&$0.87$&$0.61$&C.\\
			$ 2^{++} $&$ 7017-39i $&$90\%$&$10\%$&$0.79$&$0.79$&$0.71$&$0.56$&C.\\
			&$ 7114-4i $&$69\%$&$31\%$&$0.92$&$0.92$&$0.65$&$0.55$&C.\\
			&$ 7204-29i $&$57\%$&$43\%$&$0.94$&$0.94$&$0.85$&$0.62$&C.\\
			&$ 7276-12i $&$73\%$&$27\%$&$0.86$&$0.86$&$1.04$&$0.93$&C.\\
			\hline\hline
			
		\end{tabular}}
	\end{table*}

We further investigate the S-wave fully bottomed tetraquark~\cite{Wu:2024euj} and fully heavy tetraquark with different flavors~\cite{Wu:2024hrv} within a unified framework. A series of resonant states are predicted in $(19.7,20.0)$ GeV for the $bb\bar b\bar b$ system and $(13.2,13.5)$ GeV for the $bc\bar b\bar c$ system, which may be searched for in future experiments. All fully heavy tetraquark resonances are found to be compact tetraquark states. The fully strange tetraquarks, as cousins of the fully heavy tetraquarks, are also studied in Ref.~\cite{Ma:2024vsi}.

\subsection{Doubly heavy tetraquark}
\begin{figure}[hbt]
	\centering
	\includegraphics[width=12cm]{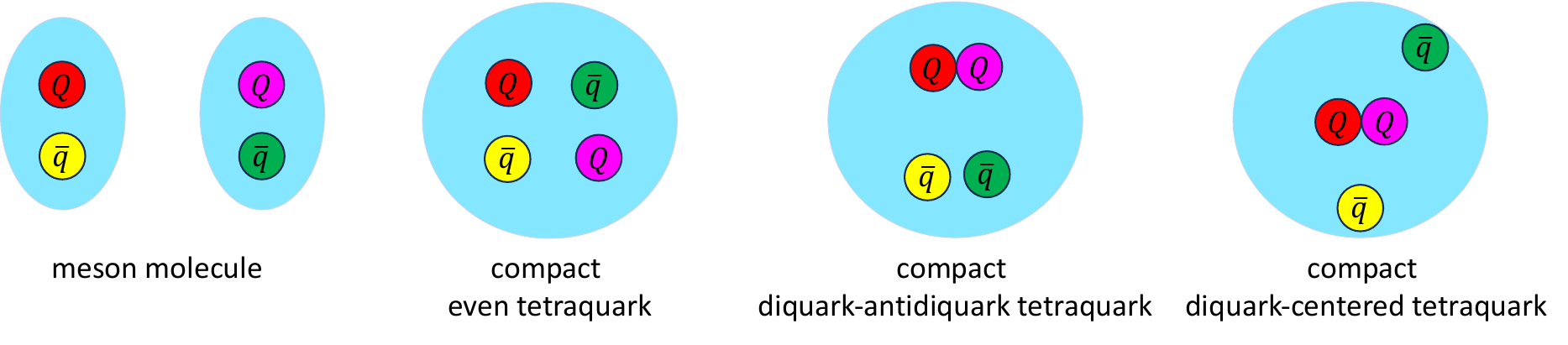}
	\caption{\scriptsize Classification of tetraquark spatial structures. The blue background indicates the confinement range.}
	\label{fig:tetra_config}
\end{figure}

As the three potential models are found to give similar results in the fully charmed tetraquark systems, we only choose the AL1 model to investigate the S-wave doubly heavy tetraquark systems $QQ^{(\prime)}\bar q\bar q$ ($Q^{\prime}=b,c$ and $q=u,d$)~\cite{Wu:2024zbx}. We observe rich spatial configurations of various doubly heavy tetraquark states by calculating their rms radii. In addition to the general classification of tetraquarks into meson molecules and compact tetraquarks, we further categorize the compact tetraquark configuration into three subtypes, as illustrated in Fig.~\ref{fig:tetra_config}.  In compact even tetraquark, the four (anti)quarks are evenly distributed, where their relative distances are of similar size. The two light antiquarks are shared by two heavy quarks, resembling the hydrogen atom, where two electrons are shared by two protons. In compact diquark-antidiquark tetraquark, diquark and antidiquark clusters are formed within the compact structure.  In compact diquark-centered tetraquark, two heavy quarks form a compact diquark core, whereas two light antiquarks orbit around the diquark. It corresponds to the QCD helium atom, where two electrons orbit around the helium nucleus. These classifications unravel different internal structures and binding mechanisms of various tetraquark states.

Some of the low-lying bound and resonant states obtained in our calculations are listed in Table~\ref{tab:QQqq}. In the $cc\bar q\bar q$ system, we obtain an isoscalar vector $D^* D$ molecular bound state with $\Delta E=-14$ MeV. Given that the model uncertainty is around tens of MeV, the bound state serves as a candidate of experimental $T_{cc}(3875)^+$. Moreover, a resonant state $T_{cc}(4031)$ is predicted near the $D^*D^*$ threshold. In the $bb\bar q\bar q$ system, we obtain two bound states with $I(J^P)=0(1^+)$, a deeply bound state with compact diquark-centered tetraquark configuration and a $\bar B^*\bar B$ molecular shallow bound state. The deeply bound state is dominated by the color configuration $\chi_{\bar{3}_c\otimes3_c}$, indicating that the strong attraction between two bottom quarks has great contributions to the binding energy. It can only decay weakly while the shallow bound state can decay radiatively to $\bar B\bar B\gamma$. Moreover, we obtain three resonant states around $10.7$ GeV, including an isoscalar $\bar B^*\bar B^*$ molecular state and two isovector compact diquark-centered tetraquarks. In the $bc\bar q\bar q$ system, we obtain three isoscalar bound states. The scalar and vector states are compact even tetraquarks with significant binding energies $\Delta E\approx-30$ MeV, while the tensor state is a weakly bound $\bar B^*D^*$ molecule with $\Delta E=-3$ MeV. In addition, we find two resonant states with $I(J^P)=0(0^+)$ and $1(2^+)$. They can be searched for in the $\bar B D$ and $\bar B^*D^*$ channels, respectively.

\begin{table*}[htbp]
	\centering
	\caption{\footnotesize The complex energies $E=M-i \Gamma / 2$ (in MeV), proportions of color configurations and spatial configurations of the doubly heavy tetraquark bound and resonant states. The binding energies $ \Delta E $ of the bound states are shown in the fourth column. The spatial configurations are shown in the last column, where C.E., C.DA., C.DC. and M. respectively indicate compact even tetraquark, compact diquark-antidiquark tetraquark, compact diquark-centered tetraquark and meson molecule. The constituent mesons are shown in the parentheses. \change{The rms radii of the tetraquarks can be found in Ref.~\cite{Wu:2024zbx}.} The "?" indicates that the rms radii are numerically unstable in the complex scaling transformation and the configuration of the state is undetermined. }
	\label{tab:QQqq}
	{\footnotesize \begin{tabular*}{\hsize}{@{}@{\extracolsep{\fill}}ccccccc@{}}
		\hline\hline
		System&$ I(J^P)$& $ M-i\Gamma/2 $ & $ \Delta E $ & $ \chi_{\bar{3}_c\otimes3_c} $ &$ \chi_{6_c\otimes \bar6_c} $ & Configuration\\
		\hline
		$ cc\bar q\bar q $&$ 0(1^+) $&$3864$ &$ -14 $& $58\%$ & $42\%$ &M.$ \,(D^*D) $\\
		&&	$ 4031-27i $&& $ 0\% $& $ 100\% $&?\\
		\hline
		$ bb\bar q\bar q $&$ 0(1^+) $&$10491$ & $ -153 $ & $97\%$ & $3\%$ & C.DC.\\
		&&$10642$ & $ -1 $ & $33\%$ & $67\%$ &M. $ \,(\bar B^*\bar B) $\\
		&&$10700-1i$ && $44\%$ & $56\%$ &M. $ \,(\bar B^*\bar B^*) $\\
		&$ 1(1^+) $
		&$10685-7i$ && $91\%$ & $9\%$ &C.DC.\\
		&$ 1(2^+) $&$10715-2i$ && $90\%$ & $10\%$&C.DC.\\
		\hline
		$ bc\bar q\bar q $&$ 0(0^+) $&$7129$ &$ -26 $& $48\%$ & $52\%$ &C.E.\\
		&&$7301-73i$ && $31\%$ & $69\%$ &C.DC.\\
		&$ 0(1^+) $&$7185$ &$ -27 $& $60\%$ & $40\%$ &C.E.\\
		&$ 0(2^+) $&$7363$ &$ -3 $& $27\%$ & $73\%$ &M. $ \,(\bar B^*D^*) $\\
		&$ 1(2^+) $&$7430-28i$ && $93\%$ & $7\%$&C.DC.\\
		\hline\hline
	\end{tabular*}}
\end{table*}
\subsection{Singly heavy tetraquark}
We further study the S-wave charm-strange tetraquark ($cs\bar q\bar q$) and bottom-strange tetraquark ($bs\bar q\bar q$) using the AL1 model~\cite{Chen:2023syh}. Table~\ref{tab:Qsqq} lists the low-lying spectra of the two systems. We identify a series of molecular bound and resonant states, including two scalar bound states ($D\bar K$ and $\bar B\bar K$), two scalar resonances ($D^*\bar K^*$ and $\bar B^*\bar K^*$), and one vector bound state ($\bar B^*\bar K$). The $D^*\bar K^*$ molecular resonance stands as a great candidate for the experimental $T^*_{\bar c\bar s0}(2870)$, whose mass and width are $2872\pm16$ MeV and $67\pm24$ MeV~\cite{ParticleDataGroup:2024cfk}. The theoretical width is expected to be smaller than the experimental value due to the neglection of the width of $\bar K^*$ in the calculations. 

In the scalar system, the $cs\bar q\bar q$ bound and resonant states share almost the same binding energies $\Delta M$ and widths with their bottom partners, which is consistent with the molecular picture. In the hadronic molecular scheme, both the interactions and the reduced masses in the $D^{(*)}\bar K^{(*)}$ and $\bar B^{(*)}\bar K^{(*)}$ systems are similar, yielding comparable binding strengths. The scalar bound states can only decay weakly, while the scalar bottom-strange tetraquark resonance may be searched for in the $\bar B\bar K$ channel.

\begin{table*}[htbp]
	\centering
	\caption{\footnotesize The complex energies $E=M-i \Gamma / 2$ (in MeV) and spatial configurations of the $cs\bar q\bar q$ and $bs\bar q\bar q$ tetraquark bound and resonant states. The notations for spatial configuration follow those of Table~\ref{tab:QQqq}. The mass differences between the molecular states and their constituent mesons are denoted by $\Delta M$. }
	\label{tab:Qsqq}
	{\footnotesize \begin{tabular*}{\hsize}{@{}@{\extracolsep{\fill}}ccccccc@{}}
		\hline\hline
        &\multicolumn{3}{c}{$cs\bar q\bar q$}&\multicolumn{3}{c}{$bs\bar q\bar q$}\\
        \hline
		$ I(J^P)$& $ M-i\Gamma/2 $ & $ \Delta E $ & Configuration&$ M-i\Gamma/2 $ & $ \Delta E $ & Configuration\\
		\hline
		$0(0^+)$&$2350$&$-3$&M.($D\bar K$)&$5781$&$-3$&M.($\bar B\bar K$)\\
		&$2906-10i$&$-14$&M.($D^*\bar K^*$)&$6240-9i$&$-14$&M.($\bar B^*\bar K^*$)\\
		$0(1^+)$&&&&$5840$&$-1$&M.($\bar B^*\bar K$)\\
		
		\hline\hline
	\end{tabular*}}
\end{table*}
\section{Summary}
The S-wave fully heavy ($QQ\bar Q\bar Q$), doubly heavy ($QQ^{(\prime)}\bar q\bar q$), and singly heavy ($Qs\bar q\bar q$) tetraquark systems are investigated within the quark potential model. The complex scaling method is used to identify bound and resonant states, and the Gaussian expansion method is used to solve the complex-scaled Schr\"odinger equation. Good candidates for experimental states in different systems are found, including fully charmed tetraquark resonances $X(6900),X(7200)$~\cite{LHCb:2020bwg,ATLAS:2023bft,CMS:2023owd}, doubly charmed tetraquark bound state $T_{cc}(3875)^+$~\cite{LHCb:2021vvq,LHCb:2021auc}, and charm-strange tetraquark resonance $T_{\bar c\bar s0}^*(2870)$~\cite{LHCb:2020bls,LHCb:2020pxc}. Moreover, a series of tetraquark states are predicted and may be searched for in near future experiments, such as $ bc\bar q\bar q, cs\bar q\bar q, bs\bar q\bar q$ bound states, $bc\bar b\bar c, cc\bar q\bar q, bc\bar q\bar q, bs\bar q\bar q$ resonances, and so on. The spatial structures of tetraquark states are identified by the rms radii. We find that all fully heavy tetraquarks have compact tetraquark configuration, all low-lying heavy-strange tetraquark are meson molecules, while doubly heavy tetraquarks exhibit rich configurations, including meson molecules, compact even tetraquarks (QCD hydrogen molecule), compact diquark-antidiquark tetraquarks, and compact diquark-centered tetraquarks (QCD helium atom). The various internal structures reveal different binding mechanisms of tetraquark states. 

\section*{ACKNOWLEDGMENTS}
This project was supported by the National
Natural Science Foundation of China (12475137). The computational resources were supported by the high-performance computing platform of Peking University.

\bibliographystyle{elsarticle-num}
\bibliography{qcd25}

\begin{thebibliography}{10}
\expandafter\ifx\csname url\endcsname\relax
  \def\url#1{\texttt{#1}}\fi
\expandafter\ifx\csname urlprefix\endcsname\relax\def\urlprefix{URL }\fi
\expandafter\ifx\csname href\endcsname\relax
  \def\href#1#2{#2} \def\path#1{#1}\fi

\bibitem{LHCb:2021vvq}
R.~Aaij, et~al., {Observation of an exotic narrow doubly charmed tetraquark},
  Nature Phys. 18~(7) (2022) 751--754.
\newblock \href {http://arxiv.org/abs/2109.01038} {\path{arXiv:2109.01038}},
  \href {https://doi.org/10.1038/s41567-022-01614-y}
  {\path{doi:10.1038/s41567-022-01614-y}}.

\bibitem{LHCb:2021auc}
R.~Aaij, et~al., {Study of the doubly charmed tetraquark $T_{cc}^{+}$}, Nature
  Commun. 13~(1) (2022) 3351.
\newblock \href {http://arxiv.org/abs/2109.01056} {\path{arXiv:2109.01056}},
  \href {https://doi.org/10.1038/s41467-022-30206-w}
  {\path{doi:10.1038/s41467-022-30206-w}}.

\bibitem{LHCb:2020bls}
R.~Aaij, et~al., {A model-independent study of resonant structure in $B^+\to
  D^+D^-K^+$ decays}, Phys. Rev. Lett. 125 (2020) 242001.
\newblock \href {http://arxiv.org/abs/2009.00025} {\path{arXiv:2009.00025}},
  \href {https://doi.org/10.1103/PhysRevLett.125.242001}
  {\path{doi:10.1103/PhysRevLett.125.242001}}.

\bibitem{LHCb:2020pxc}
R.~Aaij, et~al., {Amplitude analysis of the $B^+\to D^+D^-K^+$ decay}, Phys.
  Rev. D 102 (2020) 112003.
\newblock \href {http://arxiv.org/abs/2009.00026} {\path{arXiv:2009.00026}},
  \href {https://doi.org/10.1103/PhysRevD.102.112003}
  {\path{doi:10.1103/PhysRevD.102.112003}}.

\bibitem{LHCb:2020bwg}
R.~Aaij, et~al., {Observation of structure in the $J /\psi$ -pair mass
  spectrum}, Sci. Bull. 65~(23) (2020) 1983--1993.
\newblock \href {http://arxiv.org/abs/2006.16957} {\path{arXiv:2006.16957}},
  \href {https://doi.org/10.1016/j.scib.2020.08.032}
  {\path{doi:10.1016/j.scib.2020.08.032}}.

\bibitem{ATLAS:2023bft}
G.~Aad, et~al., {Observation of an Excess of Dicharmonium Events in the
  Four-Muon Final State with the ATLAS Detector}, Phys. Rev. Lett. 131~(15)
  (2023) 151902.
\newblock \href {http://arxiv.org/abs/2304.08962} {\path{arXiv:2304.08962}},
  \href {https://doi.org/10.1103/PhysRevLett.131.151902}
  {\path{doi:10.1103/PhysRevLett.131.151902}}.

\bibitem{CMS:2023owd}
A.~Hayrapetyan, et~al., {New Structures in the
  J/{\ensuremath{\psi}}J/{\ensuremath{\psi}} Mass Spectrum in Proton-Proton
  Collisions at s=13{\,}{\,}TeV}, Phys. Rev. Lett. 132~(11) (2024) 111901.
\newblock \href {http://arxiv.org/abs/2306.07164} {\path{arXiv:2306.07164}},
  \href {https://doi.org/10.1103/PhysRevLett.132.111901}
  {\path{doi:10.1103/PhysRevLett.132.111901}}.

\bibitem{Meng:2020knc}
Q.~Meng, E.~Hiyama, A.~Hosaka, M.~Oka, P.~Gubler, K.~U. Can, T.~T. Takahashi,
  H.~S. Zong, {Stable double-heavy tetraquarks: spectrum and structure}, Phys.
  Lett. B 814 (2021) 136095.
\newblock \href {http://arxiv.org/abs/2009.14493} {\path{arXiv:2009.14493}},
  \href {https://doi.org/10.1016/j.physletb.2021.136095}
  {\path{doi:10.1016/j.physletb.2021.136095}}.

\bibitem{Deng:2021gnb}
C.~Deng, S.-L. Zhu, {Tcc+ and its partners}, Phys. Rev. D 105~(5) (2022)
  054015.
\newblock \href {http://arxiv.org/abs/2112.12472} {\path{arXiv:2112.12472}},
  \href {https://doi.org/10.1103/PhysRevD.105.054015}
  {\path{doi:10.1103/PhysRevD.105.054015}}.

\bibitem{Meng:2021yjr}
Q.~Meng, M.~Harada, E.~Hiyama, A.~Hosaka, M.~Oka, {Doubly heavy tetraquark
  resonant states}, Phys. Lett. B 824 (2022) 136800.
\newblock \href {http://arxiv.org/abs/2106.11868} {\path{arXiv:2106.11868}},
  \href {https://doi.org/10.1016/j.physletb.2021.136800}
  {\path{doi:10.1016/j.physletb.2021.136800}}.

\bibitem{Meng:2023for}
Q.~Meng, E.~Hiyama, M.~Oka, A.~Hosaka, C.~Xu, {Doubly heavy tetraquarks
  including one-pion exchange potential}, Phys. Lett. B 846 (2023) 138221.
\newblock \href {http://arxiv.org/abs/2308.05466} {\path{arXiv:2308.05466}},
  \href {https://doi.org/10.1016/j.physletb.2023.138221}
  {\path{doi:10.1016/j.physletb.2023.138221}}.

\bibitem{Wang:2022yes}
G.-J. Wang, Q.~Meng, M.~Oka, {S-wave fully charmed tetraquark resonant states},
  Phys. Rev. D 106~(9) (2022) 096005.
\newblock \href {http://arxiv.org/abs/2208.07292} {\path{arXiv:2208.07292}},
  \href {https://doi.org/10.1103/PhysRevD.106.096005}
  {\path{doi:10.1103/PhysRevD.106.096005}}.

\bibitem{HIYAMA2003223}
E.~Hiyama, Y.~Kino, M.~Kamimura,
  \href{https://www.sciencedirect.com/science/article/pii/S0146641003900159}{Gaussian
  expansion method for few-body systems}, Progress in Particle and Nuclear
  Physics 51~(1) (2003) 223--307.
\newblock \href {https://doi.org/https://doi.org/10.1016/S0146-6410(03)90015-9}
  {\path{doi:https://doi.org/10.1016/S0146-6410(03)90015-9}}.
\newline\urlprefix\url{https://www.sciencedirect.com/science/article/pii/S0146641003900159}

\bibitem{Aguilar1971}
J.~Aguilar, J.~M. Combes, {A class of analytic perturbations for one-body
  schroedinger hamiltonians}, Commun. Math. Phys. 22 (1971) 269--279.
\newblock \href {https://doi.org/10.1007/BF01877510}
  {\path{doi:10.1007/BF01877510}}.

\bibitem{Balslev1971}
E.~Balslev, J.~M. Combes, {Spectral properties of many-body schroedinger
  operators with dilatation-analytic interactions}, Commun. Math. Phys. 22
  (1971) 280--294.
\newblock \href {https://doi.org/10.1007/BF01877511}
  {\path{doi:10.1007/BF01877511}}.

\bibitem{Aoyama2006}
S.~Aoyama, T.~Myo, K.~Kat{\=o}, K.~Ikeda, The complex scaling method for
  many-body resonances and its applications to three-body resonances, Progress
  of theoretical physics 116~(1) (2006) 1--35.

\bibitem{SilvestreBrac1996}
B.~Silvestre-Brac, {Spectrum and static properties of heavy baryons}, Few Body
  Syst. 20 (1996) 1--25.
\newblock \href {https://doi.org/10.1007/s006010050028}
  {\path{doi:10.1007/s006010050028}}.

\bibitem{Barnes:2005pb}
T.~Barnes, S.~Godfrey, E.~S. Swanson, {Higher charmonia}, Phys. Rev. D 72
  (2005) 054026.
\newblock \href {http://arxiv.org/abs/hep-ph/0505002}
  {\path{arXiv:hep-ph/0505002}}, \href
  {https://doi.org/10.1103/PhysRevD.72.054026}
  {\path{doi:10.1103/PhysRevD.72.054026}}.

\bibitem{ParticleDataGroup:2024cfk}
S.~Navas, et~al., {Review of particle physics}, Phys. Rev. D 110~(3) (2024)
  030001.
\newblock \href {https://doi.org/10.1103/PhysRevD.110.030001}
  {\path{doi:10.1103/PhysRevD.110.030001}}.

\bibitem{Wu:2024ocq}
W.-L. Wu, S.-L. Zhu, {Fully charmed P-wave tetraquark resonant states in the
  quark model}, Phys. Rev. D 111~(3) (2025) 034044.
\newblock \href {http://arxiv.org/abs/2411.17962} {\path{arXiv:2411.17962}},
  \href {https://doi.org/10.1103/PhysRevD.111.034044}
  {\path{doi:10.1103/PhysRevD.111.034044}}.

\bibitem{Wu:2024euj}
W.-L. Wu, Y.-K. Chen, L.~Meng, S.-L. Zhu, {Benchmark calculations of fully
  heavy compact and molecular tetraquark states}, Phys. Rev. D 109~(5) (2024)
  054034.
\newblock \href {http://arxiv.org/abs/2401.14899} {\path{arXiv:2401.14899}},
  \href {https://doi.org/10.1103/PhysRevD.109.054034}
  {\path{doi:10.1103/PhysRevD.109.054034}}.

\bibitem{Chen:2023syh}
Y.-K. Chen, W.-L. Wu, L.~Meng, S.-L. Zhu, {Unified description of the
  Qsq{\textasciimacron}q{\textasciimacron} molecular bound states, molecular
  resonances, and compact tetraquark states in the quark potential model},
  Phys. Rev. D 109~(1) (2024) 014010.
\newblock \href {http://arxiv.org/abs/2310.14597} {\path{arXiv:2310.14597}},
  \href {https://doi.org/10.1103/PhysRevD.109.014010}
  {\path{doi:10.1103/PhysRevD.109.014010}}.

\bibitem{Wu:2024zbx}
W.-L. Wu, Y.~Ma, Y.-K. Chen, L.~Meng, S.-L. Zhu, {Doubly heavy tetraquark bound
  and resonant states}, Phys. Rev. D 110~(9) (2024) 094041.
\newblock \href {http://arxiv.org/abs/2409.03373} {\path{arXiv:2409.03373}},
  \href {https://doi.org/10.1103/PhysRevD.110.094041}
  {\path{doi:10.1103/PhysRevD.110.094041}}.

\bibitem{CMS:2025fpt}
A.~Hayrapetyan, et~al., {Determination of the spin and parity of all-charm
  tetraquarks} (6 2025).
\newblock \href {http://arxiv.org/abs/2506.07944} {\path{arXiv:2506.07944}}.

\bibitem{Wu:2024hrv}
W.-L. Wu, Y.~Ma, Y.-K. Chen, L.~Meng, S.-L. Zhu, {Fully heavy tetraquark
  resonant states with different flavors}, Phys. Rev. D 110~(3) (2024) 034030.
\newblock \href {http://arxiv.org/abs/2406.17824} {\path{arXiv:2406.17824}},
  \href {https://doi.org/10.1103/PhysRevD.110.034030}
  {\path{doi:10.1103/PhysRevD.110.034030}}.

\bibitem{Ma:2024vsi}
Y.~Ma, W.-L. Wu, L.~Meng, Y.-K. Chen, S.-L. Zhu, {Fully strange tetraquark
  resonant states as the cousins of X(6900)}, Phys. Rev. D 110~(7) (2024)
  074026.
\newblock \href {http://arxiv.org/abs/2408.00503} {\path{arXiv:2408.00503}},
  \href {https://doi.org/10.1103/PhysRevD.110.074026}
  {\path{doi:10.1103/PhysRevD.110.074026}}.

\end{thebibliography}

\end{document}